\documentclass[12pt]{article} \hoffset=-2.15cm \textwidth=18cm  \voffset=-2.9cm \baselineskip=5cm \textheight= 24.5cm 
\newcommand{\ee}{\end{equation}}
\newcommand{\be}{\begin{equation}}
\def\ba{\begin{eqnarray}}  \def\ea{\end{eqnarray}}
  \def\hy{hyperbolic } 
  \def\t{\tau} \def\r{\rho}  
   \def\th{\theta} 
 \def\f{\phi}   
  \def\a{\alpha}  
  \def\D{\Delta}  
  
\def\ra{\rightarrow} \def\Ra{\Rightarrow} \def\2{{1\over2}}
\def\Lra{\Leftrightarrow}
\begin{document}
\title{Space-time trigonometry and
formalization of the ``Twin Paradox" for uniform and accelerated motions}
\author{Dino Boccaletti\footnote{Dipartimento di Matematica, Universit\`a di
Roma \lq\lq La Sapienza", Roma, Italy \protect \\
e-mail boccaletti@uniroma1.it}, Francesco Catoni,
\\ Vincenzo Catoni\footnote{e-mail vjncenzo@yahoo.it},
}
\date{\today}
\maketitle
{\bf Abstract} -
The formal structure of the early Einstein's Special Relativity follows
the axiomatic deductive method of Euclidean geometry. In this paper we
show the deep-rooted relation between Euclidean and space-time geometries
that are both linked to a two-dimensional number system: the complex and
hyperbolic numbers, respectively.\\
By studying the properties of these numbers together, pseudo-Euclidean trigonometry
has been formalized with an axiomatic deductive method and this allows us to give
a complete quantitative formalization of the twin paradox in a familiar ``Euclidean"
way for uniform motions as well as for accelerated ones.

\tableofcontents
\setcounter{equation}{0}
\section{Introduction}
The final part of \S \ 4 of the famous Einstein's 1905 special relativity paper \cite{1}
contains the sentences concerning moving clocks on which volumes have been written:
``.. If we assume that the result proved for a polygonal line is also valid for a
continuously curved line, we obtain the theorem: If one of two synchronous clocks at
A is moved in a closed curve with constant velocity until it returns to A, the journey
lasting $t$ seconds, then the clock that moved runs $\2\,t\left(\frac{v}{c}\right)^2$
seconds \cite{2} slower than the one that remained at rest".\\
About six years later, on 10 April 1911, at the Philosophy Congress at Bologna, Paul
Langevin replaced the clocks A and B with human observers and ``twin paradox" officially was
born. \\
Langevin, using the example of a space traveller who travels a distance L (measured by
someone at rest on the earth) in a straight line to a star in one year and than abruptly
turns around and returns on the same line, wrote: ``..Revenu \`a la Terre ayant vielli
deux ans, il sortira de son arche et trouvera notre globe vielli deux cents ans si sa
vitesse est rest\'ee dans l'interval inf\'erieure d'un vingt-milli\`eme seulement \`a la
vitesse de la lumi\`ere."\cite{3}
We must remark that Langevin, besides not rejecting ether's existence, stresses the point
which will be the subject of the subsequent discussions, that is the asymmetry between the
two reference frames. \\
The space traveller undergoes an acceleration halfway of his journey, while the twin at
rest in the earth reference frame always remains in an inertial frame. \\
For Langevin, every acceleration has an absolute meaning. Even though the effect foreseen
by Einstein's theory has got several experimental confirmations, the contribution of
accelerated stretches of the path still stands as a subject of discussion and controversies. \\
Aiming to not causing misunderstandings, we stress that the discussions we allude are
rigorously confined to the ambit of special relativity, that is to the space-time of special
relativity. \\
It is in this ambit that Rindler says: ``...If an ideal clock moves {\it nonuniformly }
through an inertial frame, we shall {\it assume }that acceleration as such has no effect
on the rate of the clock, i.e., that its instantaneous rate depends only on its
instantaneous speed $v$ according to the above rule. Unfortunately, there is no
way of {\it proving}  this.
Various effects of acceleration on a clock would be consistent with S. R. Our assumption
is one of simplicity - or it can be regarded as a definition of an ``ideal" clock.
We shall call it the clock hypothesis." \cite{4}. \\
We think that a conclusion on the role of the accelerated motions and, most of all, an
evaluation of the amount of the slowing down of an accelerated clock can only be reached
through a rigorous and exhaustive exploitation of the mathematics of special relativity. \\
If the theory has no logical inconsistencies, the theory itself must thereby provide a
completely accurate account of the asymmetrical aging process. \\
Even if Minkowski gave a geometrical interpretation of the special relativity space-time
shortly after (1907-1908) Einstein's fundamental paper, a mathematical tool exploitable
in the context of the Minkowski space-time has begun to be carried out only few decades
ago (\cite{Ya2}-\cite{vin}). \\
This mathematical tool is based on the use of hyperbolic numbers, introduced by S. Lie in
the late XIX century \cite{Lie}. \\
In analogy with the procedures applied in the case of complex numbers, it is possible to
formalize, also for the hyperbolic numbers, a space-time geometry and a trigonometry
following the same Euclidean axiomatic-deductive method \cite{vin, rob}. \\
In this paper, we first summarize the introduction of the hyperbolic space-time trigonometry
and then apply it to formalize the twin paradox for inertial motions as well as for
accelerated ones. \\
The self-consistency of the method allows us to solve any problem in the Minkowski
space-time through an elementary approach as if we were working on Euclidean plane. \\
We would conclude this introduction with an epistemological consideration which, in the
centenary of special relativity, turns out to be a recognition of the Einstein's insight. \\
We know that Einstein formalized special relativity by starting from two axioms and applying
 the axiomatic-deductive method of Euclidean geometry \cite{1}.\\
Euclid's geometry and special relativity are both associated with group theory and, in
the same way, to complex and hyperbolic numbers \cite{cz} at the extent that we
can say: Euclidean geometry and the geometry of Minkowski space-time are both deriving
from a second degree algebraic equation \cite{Ya2, geo}. \\ In particular: \\
1. from square roots of negative quantities we have complex numbers and Euclidean geometry;\\
2. from square roots of positive quantities we have hyperbolic numbers and Minkowski
space-time geometry. \\
Then these two geometries have a common source and for this reason can be considered
as equivalent. \\
Perhaps the ``ingenious intuition" of the general laws of nature let him guess an
incredible equivalence in spite of the apparent differences.
\section{Hyperbolic trigonometry}
\subsection{Basic definitions}
Complex numbers are strictly related to Euclidean geometry: indeed their
invariant (the modulus) is the same as the Pythagoric distance (Euclidean
invariant) and their unimodular multiplicative group is the Euclidean
rotation group. As it is known these properties allow us to use
complex numbers for representing plane vectors. \\
In the same way hyperbolic numbers, an extension of complex numbers \cite{Ya2, La}
defined as $$\{z=t+h\,x; \,\,h^2=1\,\,\,t,\,x \in {\bf R};\;h\notin {\bf R}\},$$
are strictly related to space-time geometry \cite{cz, vin}.
Indeed their modulus is given by (we call $\tilde z=t-h\,x$ the \hy
conjugate of $z$ as for complex numbers)
\be |z|^2=z\tilde z\equiv t^2-x^2, \label{INV}\ee
and if $t$ is given the physical meaning of a normalized time variable
(the speed of light $c=1$) and $x$ the meaning of a space variable, then
Eq. (\ref{INV}) is the Lorentz invariant of two dimensional special
relativity \cite{Fj}. Moreover their unimodular multiplicative group is the
special relativity Lorentz's group \cite{Fj, cz}. \\
Then hyperbolic numbers represent for space-time plane the same complex numbers
represent for Euclidean plane. Thanks to this correspondence and by pointing out the
analogies and the differences with these two number systems, the space-time trigonometry
has been formalized with the same rigour as Euclidean one \cite{vin}, \cite{rob}. \\
In fact the theorems of Euclidean trigonometry are usually obtained through
elementary geometry observations. Otherwise we can define in a Cartesian plane
the trigonometric functions directly from Euclid's rotation
group (as shown in the appendix \ref{inve}) and, as a consequence, the trigonometry
theorems will follow just as mathematical identities. Now since we know that the
Lorentz group enjoys, for space-time geometry, the same invariance property as
the rotation group does for Euclidean geometry, we can introduce in a Cartesian plane
the hyperbolic trigonometric functions through the properties of Lorentz group
described by hyperbolic numbers \cite{vin}. The importance of this introduction relies
on the fact that we do not have for pseudo-Euclidean geometry the same intuitive
vision we have for Euclidean geometry.
The obtained results and their complete coherence will provide a Euclidean
picture of pseudo-Euclidean geometry. \\
This picture can be considered analogous to the representation, on a Cartesian
plane, of the surface differential geometry by means of the distance between two points
given by the Lorentz invariant (space-time interval or proper time), instead of
the Pythagorean one. In this plane the
Lorentz transformations (uniform motions) are represented by straight-lines,
and the curved lines represent non-uniform (accelerated) motions \cite{cz}. In particular
the constant accelerated motions are given by an arm of equilateral hyperbola.
\cite{Na, MTW}

Here we briefly summarize some fundamental properties of hyperbolic  numbers.\\
This number system has been introduced by S. Lie \cite{Lie} as a two
dimensional example of the more general class of the commutative
hypercomplex numbers systems \cite{geo}.

Now let us introduce a hyperbolic  plane on the analogy of the Gauss-Argand plane of the
complex variable. In this plane we associate the point $P\equiv(t,\,\,x)$ to
the hyperbolic number $z=t+h\,x$.  If we represent this number on a Cartesian
plane, in this plane the square of $z$ distance ($D^2$) from the origin of the
coordinate axes is defined as
\be
D^2=z\,\tilde z\equiv t^2-x^2. \label{disq}
\ee
Let us consider the multiplicative inverse of $z$ that, if existing, is given by:
$1/z\equiv \tilde z/z\tilde z.$ This implies that $z$ does not have an
inverse when $z\tilde z\equiv t^2-x^2=0$, i.e., when $x=\pm t$, or alternatively when
$z=t\pm h\,t$. These two straight-lines in the hyperbolic plane, whose
elements have no inverses, divide the hyperbolic  plane in four sectors
 that can be called {\it Right sector (Rs), Up sector (Us), Left sector (Ls),} and
{\it Down sector (Ds)}.
This property is the same as that of the Minkowski plane
and this correspondence assigns the physical meaning of {\sf proper time} (space-time interval) to the
definition of distance \cite{Na}.
Let us now consider the quantity $t^2-x^2$, which is positive in the $Rs, Ls\;
(|t|>|x|)$ sectors, and negative in the $Us, Ds\;(|t|<|x|)$ sectors. This
quantity, as known from special relativity, must have its sign and appear
in this quadratic form. In particular, in the case we had to use the
linear form $\sqrt{t^2-x^2}$, (the modulus of hyperbolic  numbers, or the
triangle side length), we will follow the definition of Yaglom (\cite{Ya2} p. 180)
and Chabat (\cite{La} p. 51), and take the absolute value of argument
$\sqrt{|t^2-x^2|}$.

Now let us introduce the hyperbolic  exponential function and
hyperbolic polar transformation. \\
The hyperbolic  exponential function in pseudo-Euclidean geometry plays the same
important role as the complex exponential function in Euclidean geometry.
Comparing absolutely convergent series it can be written \cite{Fj, La}\\
for $|t|>|x|,\,\,t>0$ (i.e., $t,\,x\in Rs$)
\be
t+h\,x=
\exp[\r'+h\,\th]\equiv 
\exp[\r'](\cosh \th+h\,\sinh \th)     \label{exp}
\ee
The exponential function allows us to introduce the {\sf hyperbolic polar
transformation}. \\
Following \cite{Fj, La} we define the {\it  radial coordinate} as
$$\exp[\r']\Ra \r=\sqrt{t^2-x^2}$$ and {\it the angular coordinate} as
$$\th=\tanh^{-1}(x/t)\equiv \tanh^{-1}v $$
Then the hyperbolic polar  transformation is defined as
\be
t+h\,x \Lra \r \exp [h\th] \equiv  \r(\cosh\th+h\sinh\th). \label{expr}
\ee
Given two points $P_j\equiv z_j\equiv (x_j,\,y_j) ,\;P_k\equiv z_k\equiv
(x_k,\,y_k)$ we define the {\it ``square distance"} between them by extending
Eq. (\ref{disq})
\be
D_{j,\,k}=(z_j-z_k)(\tilde z_j-\tilde z_k).
\label{disq1}
\ee
{\it As a general rule we indicate the square of the segment lengths by capital
letters, and by the same small letters the square root of their absolute value}
\cite{Ya2, La}
\be
d_{j,\,k}=\sqrt{|D_{j,\,k}|}. \label{disq2}
\ee
Following the usual convention \cite{Na}, a segment or line is said to be
{\it timelike} ({\it spacelike}) if it is parallel to a line through the origin
located in the sectors containing the axis {\it t (x)}. Then the segment
$\overline{P_j P}_k$ is time-like (space-like) if $D_{j,\,k}>0$ ($D_{j,\,k}<0$),
and {\it lightlike} or null lines if $D_{j,\,k}=0$.
\subsection{Basic relations in the pseudo-Euclidean triangles}    \label{2.2}
The guide-lines for ``Euclidean" formalization of space-time trigonometry
are summarized in appendix \ref{AA}. An exhaustive treatment of this subject can
be found in \cite{vin}. Here we only report the conclusions which allow us to
formalize the twin paradox. \\
As a matter of fact the same laws that hold for Euclidean trigonometry are
true for pseudo-Euclidean trigonometry and the latter can be obtained from the
former by means of the following substitutions \cite{vin}\\
1 ) Euclidean distance $x^2+y^2\Ra$ pseudo-Euclidean distance $t^2-x^2$    \\
2 ) Circular angles $\Ra$ hyperbolic angles. \\
3 ) The straight-lines equations are expressed by means of hyperbolic
trigonometric functions \cite{pao}.\\
Moreover, all the theorems that hold in Euclidean geometry for the circle
(invariant curve \cite{Ya2}) are changed in the same ones for equilateral
hyperbolas (invariant curve for pseudo-Euclidean plane \cite{Fj}). \\
In particular, the equation of an equilateral hyperbola depends on three conditions,
that can be the same we require to determine a circle in a Cartesian plane. \\ As a
function of the center coordinates ($t_C,\;x_C$) and the diameter ($2\,p$), it is given by
\be
(t-t_C)^2-(x-x_C)^2=p^2
\ee
or, in parametric form, by Eq. (\ref{I}) of the next section.

Now we recall the theorems that will be used in this paper. We call
$\r_i$, for $i=1,\,2,\,3$, the lengths of three sides of a triangle, $\th_i$ the
opposite angles to $\r_i$, $2\,p$ the ``diameter" of the equilateral
hyperbola ``circumscribed" to the triangle; we have\\
\noindent {\it $\bullet$ Law of sines.}
\be
\frac{\r_1}{\sinh_e\th_1}=\frac{\r_2}{\sinh_e\th_2}=
\frac{\r_3}{\sinh_e\th_3}=2\,p. \label{tsen}
\ee
\noindent {\it $\bullet$ Second law of cosines.}\\ 
\be
\r_i=|\r_j\cosh_e \th_k+\r_k\cosh_e \th_j| \label{proiezioni}.
\ee
\noindent {\it $\bullet$ The sum of the internal angles in a triangle satisfies the same
relations as in Euclidean triangles} \cite{nota2}
\be \sinh_e(\th_1+\th_2+\th_3)=0,\,\,\cosh_e(\th_1+\th_2+\th_3)=- 1. \label{somma}
\ee
\noindent {\it $\bullet$ If we have points $A$ and $B$ on the same
arm of an equilateral hyperbola, for any point $P$ outside arc $AB$,
hyperbolic angles $\widehat{APB}$ are the same. If we call $C$ the center
of the equilateral hyperbola, we also have $\widehat{ACB}=2\,\widehat{APB}$.  }

From a mathematical point of view, with the extension of the trigonometric
hyperbolic functions exposed in \cite{Fj}, \cite{vin} and summarized in
appendix \ref{inv}, the hyperbolic trigonometry holds in the whole hyperbolic
plane and allows us to consider
triangles having sides in whatever direction \cite{vin}\cite{nota3}.
As far as this paper is concerned we are dealing with physical phenomena
that are represented just in (Rs) \cite{Na}, and the hyperbolic functions
are the classical ones, taking into account that the hyperbolic angles can
be measured, in particular for the parametric form of equilateral hyperbola,
with respect to a straight-line parallel to the $x$ axis. \\
Moreover, since the above summary may result inadequate to make the reader
familiar with hyperbolic trigonometry we preferred, in some examples,
not to use directly the hyperbolic counterpart of Euclidean theorems. We obtain
the results by means of simple
mathematics and afterwards show the mentioned correspondence.
\section{Mathematical formalization of the twin paradox}
As we have already emphasized in the introduction, a consequence of the Lorentz
transformations is the so
called ``Twin paradox". After a century this problem continues to be the subject
of many papers, not only relative to experimental tests \cite{CERN} but also
regarding physical and philosophical considerations \cite{arX}.\\
In this paper we want to show how the formalization of hyperbolic trigonometry \cite{vin}
allows us, with elementary mathematics, a formalization of this problem both
for uniform and accelerated motions.
\subsection{Inertial motions}
\begin{figure}[h]
\begin{center}
\setlength{\unitlength}{0.25mm}
\begin{picture}(400,250)(0,0)
\put(0,0){\framebox(400,250)}

\put(20,125){\vector(1,0){360}} 
\put(40,20){\vector(0,1){210}} 
\put(27,110){\mbox{O}} 
\put(28,215){\mbox{x}} 
\put(370,112){\mbox{t}} 

\thicklines
\put(40, 125){\line(5,2){150}}
\put(190,185){\line(5,-3){100}}
\put(40, 125){\line(2,-1){100}}
\put(140,75){\line(3,1){150}}
\put(40, 125){\vector(1,0){250}}
\put(190,187){\mbox{T}} 
\put(182,168){\rmfamily $\th_3$} 
\put(78,127){\rmfamily $\th_1$} 
\put(225,110){\rmfamily $\th_2'$} 
\put(255,127){\rmfamily $\th_2$} 
\put(78,110){\rmfamily $\th_1'$} 
\put(138,85){\rmfamily $\th_3'$} 
\put(140,61){\mbox{T'}} 
\put(291,127){\mbox{R}} 

\end{picture}
\caption{The twin paradox for uniform motions \label{motouni}}
\end{center}
\end{figure}
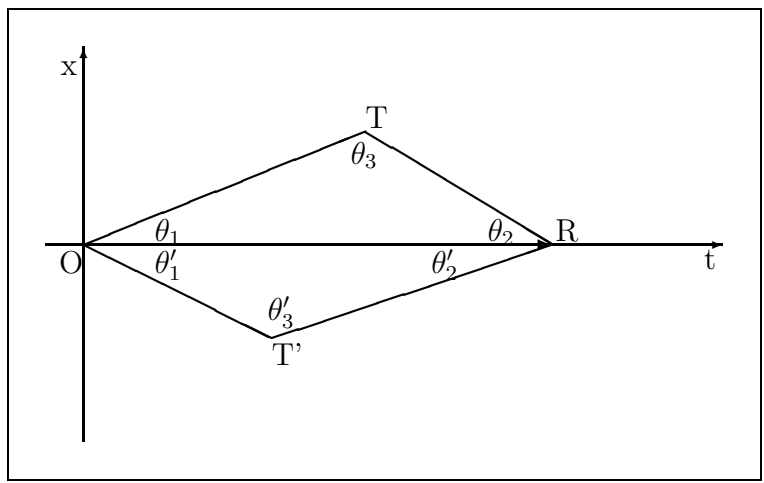

In a representative $t,\,x$ plane let us start with the following example:
a twin is steady in the point $x=0$, his path is represented by the $t$
axis. The other twin, on a rocket, starts with speed $v$ from $O\equiv(0,\,0)$
and after a time $\t_1$, at the point $T$, he reverses its direction and comes
back arriving to the point $R\equiv(\t_2,\,0)$ \cite{nota1}.
In Fig. \ref{motouni} we represent this problem by means of the triangle $OTR$. \\
From a geometrical point of view we can compare the elapsed times for the
twins by comparing the ``lengths" (proper times) of the sum
$\overline{OT}+\overline{TR}$ and of the side $\overline{OR}$. \\
The qualitative interpretation is reported in many books and is easily explained
by means of the reverse triangle inequality in space-time geometry with respect to
Euclidean geometry (\cite{dino} p. 130). Also a graphical visualization can be easily
performed considering that a segment must be reported on another by means of an equilateral
hyperbola, instead of Euclidean circle (\cite{Ya2} p. 190).

Now we will see that Euclidean formalization of space-time trigonometry \cite{vin}
allows us to obtain a simple quantitative formulation of the problem. \\
Let us call $\th_1\equiv \tanh v$ the hyperbolic angle $\widehat{ROT}$, $\th_2$
the hyperbolic angle $\widehat{ORT}$ and $\th_3$ the hyperbolic angle $\widehat{OTR}$.
From their physical meaning the angles $\th_1$ and $\th_2$ are so that the straight-lines
$\overline{OT}$ and $\overline{TR}$
are time-like \cite{Na} (in a Euclidean representation the angle of the
straight-lines with the $t$ axis must be less than $\pi/4$).\\
Let us apply to the side $\overline{OR}$ the second cosine law (\ref{proiezioni});
we have
\be\overline{OR}=\overline{OT}\,\cosh\th_1+\overline{TR}\,\cosh\th_2. \label{cos2}
\ee
It follows that the difference between the twins' proper times $\D\,\t$ is
\be
\D\,\t\equiv\overline{OR}-\overline{OT}-\overline{TR}=
\overline{OT}\,(\cosh\th_1-1)+\overline{TR} \,(\cosh\th_2-1)\label{11}.
\ee
If we call $p$ the semi-diameter of the equilateral hyperbola circumscribed to
the triangle $\widehat{OTR}$, from Eq. (\ref{tsen}) we have
 $\overline{OT}=2\,p\,\sinh\th_2;\;\overline{TR} =2\,p\,\sinh\th_1$, and
\be
\D\t=2\,p\,(\cosh\th_1\,\sinh\th_2+\cosh\th_2\,\sinh\th_1-\sinh\th_1-
\sinh\th_2) \equiv 2\,p\,[\sinh(\th_1+\th_2)-\sinh\th_1-\sinh\th_2] \label{Dt1}.
\ee
Now we can consider the following problem: given $\th_1+\th_2=const \equiv C$,
what is the relation between $\th_1$ and $\th_2$, so that
$\D\,\t$ has its greatest value?\\
The straightforward solution is
\ba
\D\t=& 2\,p\,[\sinh C-\sinh\th_1-\sinh (C-\th_1)] \qquad \qquad
 \nonumber \\
\frac{d\,(\D\t)}{d\,\th_1}\equiv &-\cosh\th_1+\cosh(C-\th_1)=0
\Ra \th_1=C/2\equiv \th_2 \label{Dt2} \\
\left.\frac{d^2\,(\D\t)}{d\,\th_1^2}\right|_{\th_1=C/2}\equiv&-\sinh(C/2)<0
\qquad\qquad\qquad\qquad\qquad\qquad   \nonumber
\ea
We have obtained the ``intuitive Euclidean" solution that the greatest difference
between the elapsed times, i.e., the shortest proper-time for the moving twin, is
obtained for $\th_1=\th_2$. For
these value Eq. (\ref{cos2}) corresponds to the well known solution \cite{1}
\be
\t_{\overline{OR}}=\t_{(\overline{OT}+\overline{TR})}\cosh \th_1\equiv
\frac{\t_{(\overline{OT}+\overline{TR})}}{\sqrt{1-v^2}}.
\ee
Now we give a geometrical interpretation of this problem. From Eq. (\ref{somma})
 we know that if $\th_1+\th_2=C$, $\th_3$ is constant too, then the
posed problem is equivalent to: what can be the position of the vertex $T$
if the starting and final points and the angle $\th_3$ are given?\\
The problem is equivalent to have, in a triangle, a side and the opposite angle. In an
equivalent problem in Euclidean geometry we know at once that the vertex $T$
does move on a circle arc. Then, from the established correspondence of circles
in Euclidean geometry to equilateral hyperbolas in pseudo-Euclidean geometry,
we have that in the present space-time problem the vertex $T$ will move on an arc of an
equilateral hyperbola.

Now let us generalize the twin paradox to the case in which both twins
change their state of motion: their motions start in $O$, both twins
move on (different) straight-lines and cross again in $R$. The graphical representation
is given by a quadrilateral figure and we call $T$ and $T'$ the other two
vertices. Since a hyperbolic rotation of the triangle does not change the
angles and the side lengths \cite{Ya2, vin},
we can rotate the figure so that the vertex $R$ lies on the $t$ axis (see
Fig. \ref{motouni}). The problem can be considered as a duplicate of the previous
one in the sense that we can compare the proper times of both twins
with the side $\overline {OR}$. If we indicate by $(')$ the quantities
referred to the triangle under the $t$ axis, we apply Eq. (\ref{Dt1}) twice
and obtain $\D\t-\D\t'$ for every specific example.\\
In particular if we have $\th_1+\th_2=\th_1'+\th_2'=C$, from the result
of Eq. (\ref{Dt2}) if follows that the youngest twin is the one for which
$\th_1$ and $\th_2$ are closer to $C/2$.
\subsection{Inertial and accelerated motions}
Now we consider some ``more realistic" examples in which uniformly accelerated
motions are taken into account.
The geometrical representation of a motion
with constant acceleration is given by an arm of an equilateral hyperbola with the
semi-diameter $p$ linked to the acceleration $a$ by the relation $p^{-1}=a$
(\cite{Na} p. 58, \cite{MTW} p. 166, \cite{noi}). \\
Obviously, the geometrical representation of a motion with non-uniform acceleration
is given by a curve which is the envelope of the equilateral hyperbolas
corresponding to the instantaneous accelerations. Or, vice versa, we can construct
in every point of a curve an ``osculating hyperbola" which has the same properties
of the osculating circle in Euclidean geometry. In fact the semi-diameter of these
hyperbolas is linked to the second derivative with respect to the line element (\cite{dino}
\S \ 3.3) as the radius of osculating circles in Euclidean geometry.\\
We also indicate by $C\equiv (t_C,\,x_C)$ its center and with $\th$ a parameter that,
from a geometrical point of view, represents a hyperbolic angle measured with respect
to an axis passing trough $C$ and parallel to $x$ axis \cite{Fj, vin}. \\
Then its equation, in parametric form, is
\be
\mbox {$\cal I$} \quad\equiv\left\{\begin{array}{l}t=t_C\pm p\,\sinh \th \\
x=x_C\pm p\,\cosh \th \end{array}\right.\quad \mbox {for }  -\infty<\th<\infty    \label{I},
\ee
where the $+$ sign refers to the upper arm of the equilateral hyperbola and the $-$ sign to
the lower one. \\
We also have
\be d\,x=\pm p\,\sinh\th\,d\,\th,\qquad d\,t=\pm p\,\cosh\th\,d\,\th \ee
and the proper time on the hyperbola
\be
\t_{\mathcal{I}}=\int_{\th_1}^{\th_2} \sqrt{d\,t^2-d\,x^2}\equiv
\int_{\th_1}^{\th_2} p\;d\,\th\equiv p\;(\th_2-\th_1) \label{proprio}   .
\ee
This relation states the link between the proper time, the acceleration, and the
hyperbolic angle and also shows that hyperbolic angles are given by the ratio
between the ``lengths" of the hyperbola arcs and the semidiameter as the
circular angles in Euclidean trigonometry are given by the ratio between circle arcs and radius.
Moreover, as in Euclidean geometry, the magnitude of hyperbolic angles is equal to twice
the area of the hyperbolic sector \cite{vin} and, taking into account that the
``area" is the same quantity in Euclidean and pseudo-Euclidean geometries, it can
be calculated in a simple Euclidean way (\cite{Ya2} p. 183). \\
In point $P$, determined by $\th=\th_1$, the velocity is given by $v\equiv
d\,x/d\,t=\tanh\th_1$
and the straight-line tangent to the hyperbola for $\th=\th_1$ is given by \cite{vin}:
\ba
x-(x_C\pm p\,\cosh\th_1)=\tanh\th_1\,[t-(t_C\pm p\,\sinh\th_1)]\Ra \nonumber \\
x\,\cosh\th_1-t\,\sinh\th_1=x_C\,\cosh\th_1-t_C\,\sinh\th_1\mp p \label{retta}
\ea
From this equation we see that $\th_1$ also represents the hyperbolic angle of
the tangent to the hyperbola with the $t$ axis. This last property means that
semi-diameter $\overline{CP}$ is pseudo-orthogonal to the tangent in $P$
(see also Fig. \ref{accI})\cite{notaper}.
This property corresponds, in Euclidean counter-part, to the well known property
of the circle where the radius is orthogonal to the tangent-line.\\
\subsubsection{First example} \label{primo}
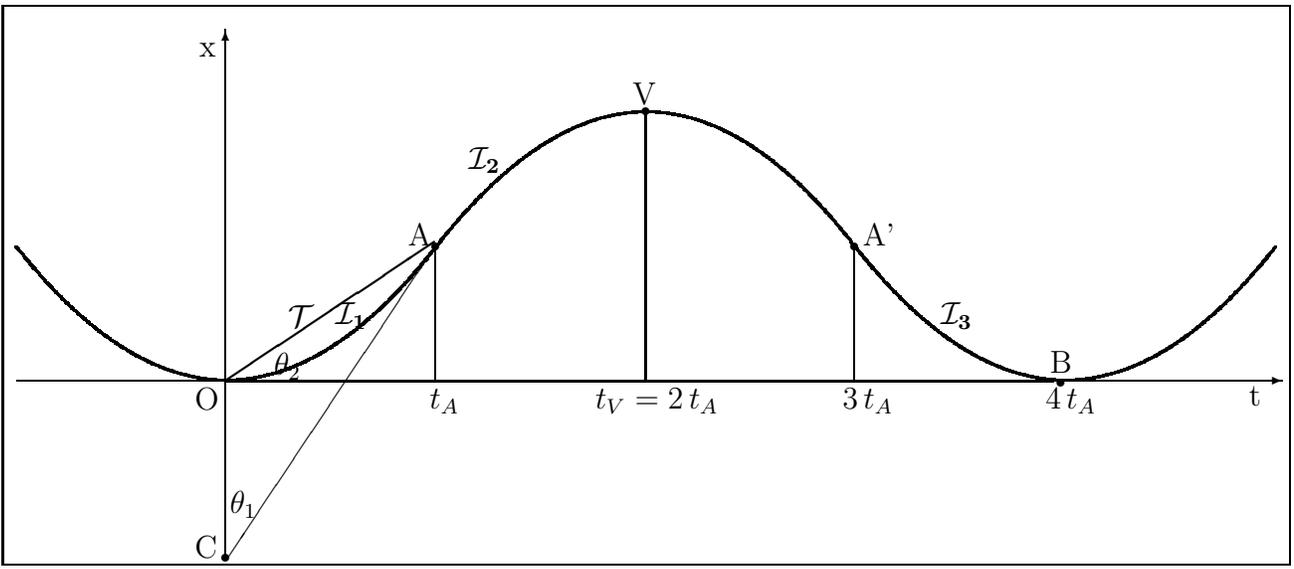
\begin{figure}[h]
\begin{center}
\setlength{\unitlength}{0.9cm}
\begin{picture}(17,6.5)(0,11.5)
\put(1.88,11.1){\vector(0,1){7.8}} 
\put(-1.2,13.71){\vector(1,0){18.7}} 
\thicklines
\qbezier(-1.22,15.69)( 1.89,11.75)( 4.98,15.69)
\qbezier( 4.98,15.69)( 8.09,19.68)(11.17,15.69)
\qbezier(11.17,15.69)(14.28,11.75)(17.38,15.69)
\put( 1.88,13.71){\line(3,2){3.09}} 
\put( 1.88,13.71){\line(1,0){12.23}} 
\thinlines
\put( 4.98,15.69){\circle*{0.12}} 
\put(1.88,11.1){\circle*{0.12}} 
\put(11.17,15.69){\circle*{0.12}} 
\put( 8.09,17.69){\circle*{0.12}} 
\put( 14.22,13.69){\circle*{0.12}} 
\put( 4.98,15.69){\line(0,-1){1.99}}
\put(11.17,15.69){\line(0,-1){1.99}}
\put( 8.09,17.69){\line(0,-1){3.98}}
\put( 4.98,15.69){\line(-2,-3){3.08}} 
\put(1.43,13.29){\rmfamily O}
\put(1.43,11.09){\rmfamily C}
\put(3.51,14.55){\rmfamily $\mathbf{\mathcal{I}_1}$}
\put(2.81,14.50){\rmfamily $\mathbf{\mathcal{T}}$}
\put(4.58,15.71){\rmfamily A}
\put(11.3,15.71){\rmfamily A'}
\put(12.45,14.55){\rmfamily $\mathbf{\mathcal{I}_3}$}
\put(7.9,17.8){\rmfamily V}
\put(5.48,16.85){\rmfamily $\mathbf{\mathcal{I}_2}$}
\put(14.06,13.83){\rmfamily B}
\put(17.0,13.34){\rmfamily t}
\put(1.5,18.5){\rmfamily x}
\put(2.6,13.8){\rmfamily $\th_2$}
\put(1.95,11.75){\rmfamily $\th_1$}
\put( 4.9,13.3){\rmfamily $t_A$}
\put( 7.35,13.3){\rmfamily $t_V=2\,t_A$}
\put(11.0,13.3){\rmfamily $3\,t_A$}
\put(14.0,13.3){\rmfamily $4\,t_A$}
\put(-1.4,11.0){\framebox(19,8.25)}
\end{picture}
\caption{The motions of example \ref{primo} \label{accII}}
\end{center}
\end{figure}
We start with the following example in which the first twin after some
accelerated motions returns to the starting point with vanishing velocity.
The problem is represented in Fig. (\ref{accII}). \\
The first twin (I) starts with a constant accelerated motion with acceleration $p^{-1}$
(indicated by $\mathcal{I}_1$) from $O$ to $A$
and then a constant decelerated ($p^{-1}$) motion up to $V$ and then accelerated with
reversed velocity up to $A'$ ($\mathcal{I}_2$), then another decelerated motion
($\mathcal{I}_3$) as $\mathcal{I}_1$ up to $B\equiv (4\,t_A,\;0)$; \\
the second twin (II) moves with  a uniform motion ($\mathcal{T}_1$) that, without loss
of generality, can be represented as stationary in the point $x=0$. \\
{\it Solution.} The equilateral hyperbola $\mathcal{I}_1$ has its center
in $C\equiv(0,\;-p)$. Then we have
\be
\mathcal{I}_1 \quad\equiv\left\{\begin{array}{l}t=p\,\sinh \th \\
x=p\,(\cosh \th -1)\end{array}\right.\quad \;\mbox {for }  0<\th<\th_1        \label{I1}.
\ee
We also have $A\equiv(p\,\sinh\th_1,\;p\,\cosh\th_1-p)$. \\
The simmetry of the problem indicates that for both twins the
total elapsed times are four times the elapsed times of the first motion. \\
The proper time of twin I is obtained from Eq. (\ref{proprio})
\be
\t_I\equiv 4\,\t_{\mathcal{I}_1}=4\,\int_0^{\th_1} \sqrt{d\,t^2-d\,x^2}\equiv
4\,\int_0^{\th_1} p\;d\,\th\equiv 4\,p\,\th_1 ,\label{motohy}
\ee
the proper time of twin II is
\be
\t_{II}\equiv 4\,t_A=4\,p\,\sinh\th_1.
\ee
The difference between the elapsed times is $\D\t=4\,p\,(\sinh\th_1-\th_1)$, and
their ratio is
\be
\frac{\t_I}{\t_{II}}=\frac{\th_1}{\sinh\th_1} \label{ratio}.
\ee
For $\th_1\equiv \tanh^{-1} v\ll 1$ \cite{notag} we have $\D\t\simeq 0$, and
for $\th\gg 1\Ra \sinh\th\propto \exp[\th]$:
{\it The proper time for the accelerated motions is linear in $\th$ and the
stationary (inertial) is exponential in $\th$}. \\
Now we show that the same relation between uniform and accelerated motion holds
if we compare the motion on the side $\overline{OA}$ with the motion on hyperbola
$\mathcal{I}_1$,  and this allows us to give a simple ``Euclidean" interpretation. \\
Let us call $\th_2$ the hyperbolic angle between straight-line $OA$ and $t$
axis; the equation of straight-line $OA$ is
\be
\mathcal{T}\equiv\{x=t\,\tanh\th_2\} \label{retta1}
\ee
and we calculate $\th_2$ imposing that this straight-line crosses the hyperbola of Eq.
(\ref{I1}) for $\th=\th_1$. By substituting Eq. (\ref{retta1}) in Eq. (\ref{I1}),
we have
\be
\left\{\begin{array}{l}t=p\,\sinh \th_1 \\
t\,\tanh\th_2=p\,(\cosh \th_1 -1)\end{array}\right. \Ra \frac{\sinh\th_2}{\cosh\th_2}=
\frac{\cosh\th_1-1}{\sinh\th_1} \Ra \th_1=2\,\th_2,
\ee
 i.e., the central angle is twice the hyperbola angle on the same chord \cite{vin}.
Then we have
\be
\overline{OA}=\frac{\overline{Ot}_A}{\cosh\th_2}\equiv\frac{p\,\sinh\th_1}{\cosh\th_2}
\equiv 2\,p\,\sinh\th_2
\ee
and taking into account the proper time on the hyperbola (Eq. \ref{proprio}), we obtain
\be
\frac{\t_{\mathcal{I}}}{\t_{\mathcal{T}}}=\frac{\th_2}{\sinh\th_2}. \label{ratio2}
\ee
This relation is a general one and it is not surprising since it derives from the
correspondence (see sec. \ref{2.2}) between Euclidean and pseudo-Euclidean geometries.
In Euclidean geometry it represents the ratio between the length of a circle arc and
its chord.
\subsubsection{Second example} \label{terzo}
\begin{figure}[h]
\begin{center}
\setlength{\unitlength}{0.9cm}
\begin{picture}(10,6.5)(1.5,12.5)
\put(1.88,12.5){\vector(0,1){6.5}} 
\put(+1.2,13.71){\vector(1,0){10.5}} 
\thicklines
\qbezier( 3.20,14.8)( 6.31,17.368)(9.39,14.8)
\put(1.88,13.71){\line(5,4){1.3}} 
\put(9.39,14.8){\line(5,-4){1.32}} 
\put(1.88,13.71){\line(1,0){8.8}} 
\thinlines
\put( 3.2,14.75){\circle*{0.12}} 
\put( 3.2,14.8){\line(4,5){3.11}} 
\put(9.39,14.8){\circle*{0.12}} 
\put( 6.31,16.07){\circle*{0.12}} 
\put( 6.31,18.7){\circle*{0.12}} 
\put( 10.75,13.69){\circle*{0.12}} 
\put( 3.20,14.8){\line(0,-1){1.1}} 
\put(9.39,14.8){\line(0,-1){1.1}} 
\put( 6.31,18.7){\line(0,-1){5.0}} 
\put(1.43,13.29){\rmfamily O}
\put(2.2,14.4){\rmfamily $\mathbf{\mathcal{T}_1}$}
\put(9.95,14.4){\rmfamily $\mathbf{\mathcal{T}_1'}$}
\put(3.0,15.0){\rmfamily P}
\put(9.4,15.0){\rmfamily P'}
\put(6.45,16.2){\rmfamily V}
\put(8.52,15.6){\rmfamily $\mathbf{\mathcal{I}}$}
\put(10.7,13.83){\rmfamily R}
\put(11.5,13.34){\rmfamily t}
\put(1.5,18.5){\rmfamily x}
\put( 3.12,13.3){\rmfamily $t_P$}
\put( 5.67,13.3){\rmfamily $t_V=t_C$}
\put(9.22,13.3){\rmfamily $t_{P'}$}
\put(10.6,13.3){\rmfamily $t_R$}
\put( 2.6,13.9){\rmfamily $\th_1$} 
\put( 6.45,18.78){\rmfamily $C$}
\put(5.9,17.8){\rmfamily $\th_1$}  
\put(+1.0,12.25){\framebox(11,7.0)}

\end{picture}
\caption{The 
motions of example \ref{terzo} \label{accI}}
\end{center}
\end{figure}
Now we consider a problem that allows us to connect the two sides of the triangle
of Fig. (\ref{motouni}) by means of an equilateral hyperbola, i.e., to consider
the decelerated and accelerated motions too.

Twin I moves from $O\equiv(0,\,0)$ to $P\equiv(p\,\sinh\th_1,\;p\cosh\th_1-p)$
with a uniform motion, indicated as $\mathcal{T}_1$, then goes on with a constant
decelerated motion up to $V$ and then accelerates with reversed
velocity up to $P'$, where he has the same velocity as the initial one, and moves again
with uniform velocity up to $R\equiv(t_R,\,0)$ ($\mathcal{T}_1'$). \\
The second twin (II) moves with  a uniform motion ($\mathcal{T}_2$) which, without loss
of  generality, can be represented as stationary in the point $x=0$. \\
{\it Solution.} A mathematical formalization can be the following: let us consider the
decelerated and accelerated motions that can be represented by the equilateral
hyperbola of Eq. (\ref{I}) for $-\th_1<\th<\th_1$
and the tangent to the hyperbola for $\th=\th_1$ as given by Eq. (\ref{retta}).
This straight-line represents the motion $\mathcal{T}_1$ if it passes trough $O$.
This happens if center of the hyperbola $C\equiv(x_C,\,t_C)$ lies on the straight-line
$x_C\,\cosh\th_1-t_C\,\sinh\th_1-p=0$, where $t_C$ is given by Eq. (\ref{I}):
$t_C=t_P+p\,\sinh\th_1$. If we write down straight-line (\ref{retta}) in parametric form
\be
\mbox {$\cal T$}_1 \quad\equiv\left\{\begin{array}{l}t=\,\t\,\cosh \th_1 \\
x=\,\t\,\sinh \th_1 \end{array}\right. \label{T1} ,
\ee
where $\t$ is the proper time on the straight-line, we have at the end of the
uniform motion $P\equiv(t_P,\,x_P)$, with $t_P=\,\t\,\cosh \th_1$. \\
Then the proper time for twin I is $\t_{\overline{OP}}=\t,$
and from $P$ to the vertex of the hyperbola $\t_{\cal I}=p\,\th_1$.
The proper times of the other lines are a duplicate of these ones. \\
For twin II we have: $\t_{II}\equiv 2\,t_C=
2\,(\t\,\cosh\th_1+p\,\sinh\th_1)$. \\
Then we have
\be
\D\,\t=2\,[\t\,(\cosh\th_1-1)+p\,(\sinh\th_1-\th_1)].
\ee
The proper time on this rounded off triangle is greater than the one on the triangle,
as we shall better see in the next example. \\
The physical interpretation is that the velocity on the hyperbola arc is less
than the one on straight-lines $OT,\;TR$ of Fig. (\ref{motouni}).
From a geometrical point of view, it is a consequence
of the reverse triangle inequality or, in a more general way, we can say that the geodesic
lines between two given points (straight-lines) are the longest lines.
\subsubsection{Third example} \label{quarto}
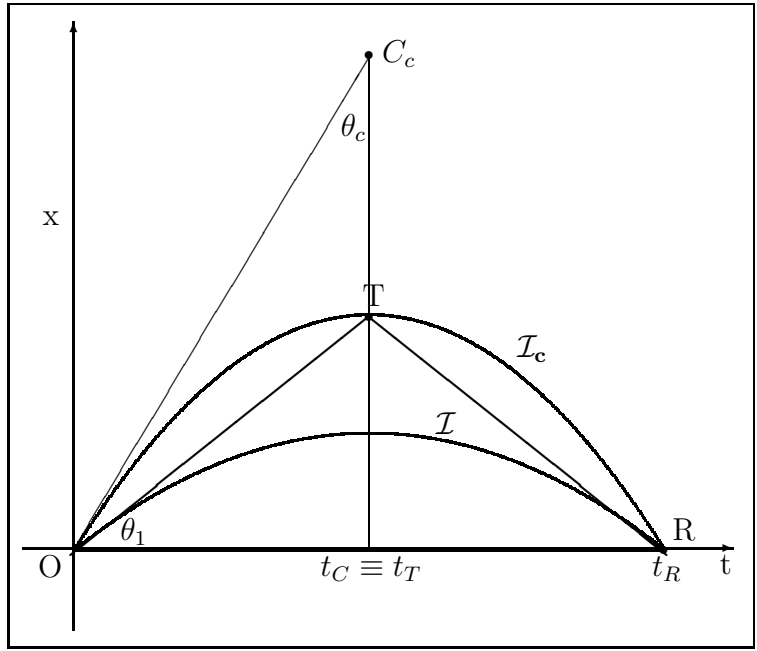
\begin{figure}[h]
\begin{center}
\setlength{\unitlength}{0.9cm}
\begin{picture}(10,9.0)(1.5,12.5)
\put(1.95,12.5){\vector(0,1){9.0}} 
\put(+1.2,13.71){\vector(1,0){10.5}} 
\thicklines
\put( 6.31,17.14){\line(-5,-4){4.4}} 
\put( 6.31,17.14){\line(5,-4){4.4}}  
\put( 1.95,13.69){\line(1,0){8.7}} 
\qbezier( 1.95,13.69)( 6.31,17.14)( 10.67,13.69)
\qbezier( 1.95,13.69)( 6.31,20.64)( 10.67,13.69)
\thinlines
\put( 6.31,17.14){\circle*{0.12}} 
\put( 6.31,21.0){\circle*{0.12}} 
\put( 10.67,13.69){\circle*{0.12}} 
\put( 6.31,21.0){\line(0,-1){7.32}} 
\put( 1.95,13.69){\line(3,5){4.4}} 
\put(1.43,13.29){\rmfamily O}
\put(7.31,15.4){\rmfamily $\mathbf{\mathcal{I}}$}
\put( 8.51,16.54){\rmfamily $\mathbf{\mathcal{I}_c}$}
\put( 6.22,17.3){\rmfamily T}
\put( 6.51,20.9){\rmfamily $C_c$}
\put(10.8,13.83){\rmfamily R}
\put(2.65,13.83){\rmfamily $\th_1$}
\put(5.9,19.8){\rmfamily $\th_c$}
\put(11.5,13.34){\rmfamily t}
\put(1.5,18.5){\rmfamily x}
\put( 5.6,13.3){\rmfamily $t_C\equiv t_T$}
\put(10.5,13.3){\rmfamily $t_R$}
\put(+1.0,12.25){\framebox(11,9.5)}
\end{picture}
\caption{The 
motions of example \ref{quarto} \label{accIII}}
\end{center}
\end{figure}
In the following example we consider the motion on the upper triangle of Fig.
\ref{motouni} with sides $\overline{OT}=\overline{TR}$ and on the
following equilateral hyperbolas \\ 1) $\mathcal{I}$ tangent in $O$ and in $R$
to sides $\overline{OT}$ and $\overline{TR}$, respectively \\
2) $\mathcal{I}_c$ circumscribed to triangle $\stackrel{\triangle}{OTR}$.

In this example we can also note a formalization of the reverse triangle inequality
(\cite{dino} p. 130). In fact, we shall see that as shorter the lines (trajectories) are
in a Euclidean representation, so longer they are in the space-time geometry.

{\it Solution.} Side $\overline{OT}$ lies on the straight-line represented by
the equation
\be x\,\cosh\th_1-t\,\sinh\th_1=0 .\label{re3}\ee
Hyperbola $\mathcal{I}$ is obtained requiring that it is tangent to
straight-line (\ref{re3}) in $O$. We obtain from Eqs. (\ref{I}, \ref{retta})
$t_C=p\,\sinh\th_1,\;x_C=p\,\cosh\th_1$ and, from the definitions of hyperbolic
trigonometry, $\overline{OT}=t_C/\cosh\th_1\equiv p\,\tanh\th_1$. \\
Let us consider hyperbola $\mathcal{I}_c$. Its vertex is $T$ and its
semi-diameter is given by Eq. (\ref{tsen}): $p_c=\overline{OT}/(2\,\sinh\th_1)
\equiv p/(2\,\cosh\th_1)$. If we call $C_c$ its center and $2\,\th_c$ angle
$\widehat{OC_cR}$, we note that $\th_c$ is a central angle of chord
$\overline{OT}$ while $\th_1$ is an hyperbola angle on equal chord
$\overline{TR}$. Then, as it has been shown in example \ref{primo}, we have
$\th_c=2\,\th_1$.\\
Now let us calculate the lengths (proper times) for the motions. \\
As to the hyperbolas, from Eq. (\ref{proprio}), we have: \\
the length of arc of hyperbola $\mathcal{I}$ between $O$ and $R$ is given
by $\t_{\mathcal{I}}=2\,p\,\th_1$, \\
the lenght of arc of $\mathcal{I}_c$ from $O$ and $R$
is given by $\t_{\mathcal{I}_c}=2\,p_c\,\th_c\equiv 2\,p\,\th_1/\cosh\th_1$. \\
For the lenghts of the segments we have: \\
$\overline{OR}\equiv 2\,t_T=2\,p\,\sinh\th_1$, and 
from Eq. (\ref{re3}) it follows $T\equiv(p\,\sinh\th_1,\,p\,\sinh\th_1\,\tanh\th_1)$, so
$\overline{OT}=\overline{TR}=p\,\tanh\th_1$. \\
Then we have the following relations:
\be \overline{OR}\equiv 2\,p\,\sinh\th_1> \mbox{ arc}(\mathcal{I})\equiv 2\,p\,\th_1>
\overline{OT}+\overline{TR}\equiv 2\,p\,\tanh\th_1>\mbox{ arc}(\mathcal{I}_c)
\equiv 2\,p\,\th_1/\cosh\th_1. \ee
We also observe that $\overline{OR}$ is a chord of $\mathcal{I}$,
$\overline{OT}$ is a chord of $\mathcal{I}_c$ and their ratios are the one given by
Eq. (\ref{ratio2}): \be
\frac{\t_{\overline{OR}}}{\t_{\mathcal{I}}}=\frac{\t_{\overline{OT}}}{\t_{\mathcal{I}_c}}
\equiv \frac{\sinh\th_1}{\th_1}. \ee
As a corollary of this example we consider the following one: given
side $\overline{OR}=\t$ (proper time of the stationary twin) what is the proper time
of twin I moving on an equilateral hyperbola, as a function of rocket
acceleration $p^{-1}$?  \\ {\it Solution.} From hyperbolic motion of Eq. (\ref{I}) we
have $t=\t/2-p\,\sinh\th$ and for $t=0$ we obtain $\th_1\Ra 2\,p\,\sinh\th_1=\t$,
and for relativistic motions ($\th_1\gg 1$) we obtain
\be \t\simeq p\,\exp[\th_1]\Ra \th_1\simeq\ln\frac{\t}{p}. \ee
Then from relation (\ref{proprio}) $\t_{\mathcal{I}}\equiv 2\,p\,\th_1=2\,p\,\ln[\t/p]
\stackrel{p\ra 0}{\longrightarrow}0$. \\
As acceleration $p^{-1}$ does increase, proper time $\t_{\mathcal{I}}$ can be
as less as we want (\cite{MTW} p. 167).
\subsubsection{Fourth example} \label{sesto}
We conclude with a more general example in which both
twins have a uniform and accelerated motion.\\ First twin (I) starts with a constant
accelerated motion and then goes on with a uniform motion,  \\ second twin (II) starts
with a uniform motion and then goes on with a constant accelerated motion. \\
{\it Solution.} We can represent this problem in the $t,\,x$ plane in the following way:\\
I starts from point $O\equiv (0,\,0)$ with an acceleration given by $p^{-1}$ ($\mathcal{I}_1$)
up to point $A\equiv(p\,\sinh\th_1;\;p\,\cosh\th_1-p)$, then goes on with a uniform motion
($\mathcal{T}_1$) up to time $t_3$ (point $B$).\\
II starts from point $O\equiv (0,\,0)$ with a uniform motion ($\mathcal{T}_2$),
(stationary in the point $x=0$) up to point $C$, in a time $t_2=\a\,p\,\sinh \th_1$
that we have written proportional to $t_A$. Then goes on with an accelerated motion
($\mathcal{I}_2$), with the same
acceleration $p^{-1}$  up to crossing the trajectory of I at time $t_3$. \\
The analytical representation of $\mathcal{I}_1$ is given by Eq. (\ref{I1}).
$\mathcal{I}_2$ is represented by
\be
 \mathcal{I}_2\quad \equiv \left\{\begin{array}{l}t=p\,(\a\,\sinh\th_1+
\sinh \th) \\
 x=p\,(\cosh \th-1)\end{array}\right. \mbox {for }  0<\th<\th_2    \label{I4},
\ee
where $\th_2$ represents the value of the hyperbolic angle in crossing point $B$
between $ \mathcal{I}_2$ and $ \mathcal{T}_1$. \\
$\mathcal{T}_1$ is given by the straight-line tangent to ${\cal I}_1$ in $\th_1$:
\be {\mathcal T}_1\equiv \{x\,\cosh\th_1-t\,\sinh\th_1=p\,(1-\cosh\th_1)\}. \label{T2} \ee
From Eqs. (\ref{I4}) and (\ref{T2}) we calculate the crossing point between
${\mathcal I}_2$ and ${\mathcal T}_1$. We have\cite{nota27}
\be \cosh(\th_2-\th_1)=\a\,\sinh^2\th_1+1 \label{th2}.\ee
Let us calculate the proper times. \\
The proper times relative to the accelerated motions are obtained from Eq. (\ref{proprio}):
$\t_{\mathcal{I}_1}=p\,\th_1,\;\t_{\mathcal{I}_2}=p\,\th_2$.
The proper time relative to $\mathcal{T}_2$ is given by $t_2=\a\,p\,\sinh\th_1$.
On straight-line ${\mathcal T}_1$, 
between points $A$ 
and $B\equiv (p\,\a\,\sinh\th_1+p\,\sinh\th_2,\;p\,\cosh\th_2-p)$,
the proper time is obtained by means of hyperbolic trigonometry \cite{vin}
\be \t_{\mathcal{T}_1}\equiv \overline{AB}=
(x_B-x_A)/\sinh\th_1\equiv p\,(\cosh\th_2-\cosh\th_1)/\sinh\th_1. \label{tauno}
\ee
Then the complete proper-times of the twins are
\be
\t_I=p\,[\th_1+(\cosh\th_2-\cosh\th_1)/\sinh\th_1],\qquad\t_{II}=
p\,(\a\,\sinh\th_1+\th_2)                                     \label{taui}.
\ee
Let us consider relativistic velocities ($v=\tanh \th_i\simeq 1\Ra \th_1,
\th_2\gg 1$); in this case we can approximate the
hyperbolic functions in Eqs. (\ref{th2}, \ref{taui}) with the
positive exponential term and, for $\a\neq 0$, we obtain from Eq. (\ref{th2}):
$\exp[\th_2-\th_1]\simeq \a\,\exp[2\,\th_1]/2$, and from Eqs. (\ref{taui})
\be
\t_I\simeq p\,(\th_1+\exp[\th_2-\th_1])\simeq p\,(\th_1+\a\exp[2\,\th_1]/2),
\qquad\t_{II}\simeq p\,(\a\,\exp[\th_1]/2+\th_2).
\ee
The greatest contributions to the proper times are given by the exponential terms
that derive from the uniform motions. If we neglect the linear terms with respect
to the exponential ones, we obtain a ratio of the proper times {\it independent
of the $\a\neq 0$ value}
\be
\t_{I}\simeq \t_{II}\,\exp[\th_1] \label{rapptau}
\ee
The twin that moves for a shorter time with uniform motion has the shortest
proper time\cite{nota28}.\\
A simplified version (an inertial and an accelerated motion between two points) is
given in \cite{MTW} (exercise 6.3 p.167) and it is considered just as
a consequence of the reverse triangle inequality.\\ With regard to the result of this
example we could ask: how is it possible that a uniform motion
close to a light-line is the longer one? We can answer this question by a glance at
Eq. (\ref{tauno}).
In fact in this equation the denominator $\sinh\th_1\gg 1$ takes into account that the
motion is close to a light-line, but in the numerator $\cosh\th_2\gg \cosh\th_1$
indicates that crossing point $B$ is so far that its contribution is the
determining term of the result we have obtained.\\
\section{Conclusions}
As we know, the twin paradox spread far and wide having been considered the
most striking exemplification of the space-time ``strangeness" of
Einstein's theory of special relativity.
What we have striven to show is that hyperbolic trigonometry supplies us with
an easy tool by which one can deal with any kinematic problem in the context
of special relativity. \\
In fact, if we consider Einstein's theory of special relativity as a
logical-deductive construction based on the two postulates of the constancy
of light's velocity and the equivalence of all inertial reference frames for
establishing physical laws, the hyperbolic space-time is the right  mathematical
structure inside which any problem must be dealt with. \\
As we have seen, the use of hyperbolic trigonometry allows us to obtain the
{\it quantitative solution} of any problem and dispels all doubts regarding the role
of acceleration in the flow of time. \\
Finally, we remark that the application of hyperbolic trigonometry to relativistic
space-time results to be a ``Euclidean way" of dealing with pseudo-Euclidean spaces.
\appendix
\section{The formalization of Euclidean and pseudo-Euclidean tri\-go\-no\-me\-tries
by means of complex or hyperbolic numbers} \label{AA}
For greater convenience of the reader we report a short exposition of paper \cite{vin}
\subsection{Rotation invariants in Euclidean plane} \label{inve}
Euclid's geometry studies the figure properties that do not depend
on their position in a plane. If these figures are represented in a Cartesian
plane we can say, in group language, that Euclid's geometry studies the
invariant properties by coordinate axes roto-translations.
It is well known that these properties can be expressed by complex numbers.
Let us consider Gauss-Argand's complex plane where a vector is represented
by $v=x+i\,y$. The axes rotation of an angle $\a$ transforms this vector in the
new vector $v'\equiv v\exp [i\a]$. Therefore we can promptly verify that the
quantity (as it is usually done we call $\bar v=x-i\,y$) $|v'|^2 \equiv v'\bar
{v'}= {v\exp [i\a]\bar v\exp [-i\a]} \equiv |v|^2$ is invariant by axes rotation.
In a similar way we find two invariant quantities related to any couple of
vectors.\cite{notainv} \\
If we consider two vectors: $v_1=x_1+iy_1$, $v_2=x_2+iy_2$; we have that
{\it the real and the imaginary part of the product $v_2 \bar v_1$ are
invariant by axes rotation.}  In fact $v^{'}_2\bar v_1^{'}= v_2\exp
[i\a]\bar v_1\exp [-i\a]\equiv v_2\,\bar{v}_1$.
Now we will see that these two invariant quantities allow an operative definition of
trigonometric functions. Let us represent the two vectors in polar coordinates:
$v_1\equiv \r_1\exp[i\f_1],\,v_2\equiv \r_2\exp[i\f_2]$. Consequently we have:
\be
v_2\bar v_1=\r_1\r_2\exp[i(\f_2-\f_1)]\equiv
\r_1\r_2[\cos(\f_2-\f_1)+i\sin(\f_2-\f_1)]. \label{polc}
\ee
As it is well known the resulting real part of this product represents
the scalar product, while the ``imaginary" part represents the modulus of the
vector product, i.e., the area of the parallelogram defined by the two vectors. \\
The two invariant quantities of Eq. (\ref{polc}) allow an operative definition
of trigonometric functions. In fact in Cartesian coordinates we have:
\be
v_2\,\bar v_1=(x_2+i\,y_2) (x_1-i\,y_1)\equiv x_1\,x_2+y_1\,y_2+i(x_1\,y_2-x_2\,y_1),
\label{carc}
\ee
and by using Eqs. (\ref{polc}) and (\ref{carc}) we obtain:
\be
\cos(\f_2-\f_1)=\frac {x_1\,x_2+y_1\,y_2}{\r_1
\,\r_2};\;\;\;\;\;\;\;\;\;\sin(\f_2-\f_1)=\frac{x_1\,y_2-x_2\,y_1}{\r_1\,\r_2}
\label{cir}
\ee
We know that the theorems of Euclid's trigonometry are usually obtained by
following a geometric approach. Now by using the Cartesian representation of
trigonometric functions given by Eqs. (\ref{cir}), it is straightforward to
control that the
trigonometry theorems are simple identities. In fact let us define a triangle
in a Cartesian plane by its vertices $P_n\equiv(x_n,\,y_n)$: from the
coordinates of these point we obtain the side lengths and, from Eqs.
(\ref{cir}) the trigonometric functions. By these definitions it is
easy to control the identities defined by the trigonometry theorems.
\subsection{Hyperbolic rotation invariants in pseudo-Euclidean plane} \label{inv}
By analogy with Euclid's trigonometry approach summarised in appendix
(\ref{inve}), we can say that {\it pseudo-Euclidean plane geometry}
studies the properties that are invariant by Lorentz
transformations (Lorentz-Poincar\`e group of special relativity) corresponding
to hyperbolic rotation as exposed in \cite{cz}.
We show afterwards, how these properties can be represented by hyperbolic numbers.

Let us define in the hyperbolic plane a hyperbolic  vector from the origin to the
point $P\equiv(t,\,x)$, as $v=t+h\,x$ and consider a hyperbolic  rotation of an
angle $\th$ that, from a physical point of view, means a Lorentz transformation
with a velocity given by $V=\tanh^{-1} \th$ \cite{Fj, Na}. From this transformation
the vector $v$ become $v'\equiv v\exp [h\,\th]$.
Therefore  we can readily verify that the quantity:
\be
|v'|^2 \equiv v'\tilde v'= {v\exp [h\,\th]\tilde v\exp
[-h\,\th]} \equiv |v|^2 \label{modh}
\ee
is invariant for hyperbolic  rotation. In a similar way we can find two
invariants related to any couple of vectors.
Let us consider two vectors $v_1=t_1+h\,x_1$ and $v_2=t_2+h\,x_2$: we have
that {\it the real and the ``hyperbolic" parts of the product $v_2 \tilde v_1$
are invariant by hyperbolic   rotation.} \\
In fact $v_2'\tilde v_1'= v_2\exp [h\a]\tilde v_1\exp [-h\a]\equiv
v_2\,\tilde{v}_1$. These two invariants allow an operative definition of the
hyperbolic trigonometric functions. To show this
let us suppose that $|t_1|>|x_1|,\;\, |t_2|>|x_2|$ and $t_1,\,t_2>0$, and
let us represent the two vectors in hyperbolic polar form (\ref{expr}): $v_1=\r_1\exp [h\,\th_1],\,v_2=\r_2\exp [h\,\th_2]$. Consequently we have
\be
  v_2\bar v_1\equiv \r_1\,\r_2 \exp[h(\th_2-\th_1)]
  \equiv \r_1\,\r_2 [\cosh (\th_2-\th_1)+h\sinh (\th_2-\th_1)]. \label{invpol}
\ee
As shown in appendix (\ref{inve}), for Euclidean plane the real part of
the vector product represents the scalar product, while the imaginary part
represents the area of the parallelogram defined by the two vectors.
In pseudo-Euclidean plane, as we know from differential geometry \cite{ei},
the real part is still the scalar product; as far as the hyperbolic part is
concerned, we see in subsection (\ref{trigpe}) that
it can be considered as a {\it pseudo-Euclidean area} \cite{Ya2}. \\
In Cartesian coordinates we have:
\be
v_2\,\tilde v_1=(t_2+h\,x_2)(t_1-h\,x_1)\equiv t_1\,t_2-x_1\,x_2
+h(t_1\,x_2-t_2\,x_1).\label{invcar}
\ee
By using Eqs. (\ref{invpol}) and (\ref{invcar}) we obtain:
\ba
\cosh(\th_2-\th_1)=\frac {t_1\,t_2-x_1\,x_2}{\r_1\,\r_2}\equiv\frac
{t_1\,t_2-x_1\,x_2}{
\sqrt{|(t^2_2-x^2_2)|\cdot|(t^2_1-x^2_1)|}} \label{cos} \\
\sinh(\th_2-\th_1)=\frac{t_1\,x_2-t_2\,x_1}{\r_1\,\r_2}
\equiv \frac{t_1\,x_2-t_2\,x_1}{ \sqrt{|(t^2_2-x^2_2)|\cdot|(t^2_1-x^2_1)|}}
\label{sen}.
\ea
If we put  $v_1\equiv (1;\,0)$ and $\th_2,\,t_2,\,x_2 \ra \th,\,t,\,x$ then
Eqs. (\ref{cos}), (\ref{sen}) can be rewritten in the form:
\be \cosh \th=\frac{t}{\sqrt{|t^2-x^2|}}; \;\;\;\;\;\;\;\; \;\;\;\;\;\;\;\; 
\sinh \th=\frac{x}{ \sqrt{|t^2-x^2|}}     \label{sen1}. \ee
The classic hyperbolic  functions are defined for $t,\,x \in Rs$. Now we can observe
that expressions in Eq. (\ref{sen1}) are valid for
$\{t,\,x\in {\bf R}| t\neq\pm x\}$ so they allow  to extend the hyperbolic  functions
in the complete $t,\,x$ plane. This extension is the same as that already
proposed in \cite{Fj}. 
These {\sf extended hyperbolic functions} have been denoted with $\cosh_e ,\,
\sinh_e$ in the paper \cite{vin}. In tab. (\ref{tab2}) the relations between
$\cosh_e ,\,\sinh_e$ and traditional hyperbolic  functions are reported.

\begin{table}[h] \begin{center} \caption{Relations between functions $\cosh_e,\,\sinh_e$ obtained from Eq.
(\ref{sen1}) and classic \hy  functions. The \hy angle $\th$ in the last four
columns is calculated referring to semi- axes $t,\,-t,\,x,\,-x$, respectively.} \label{tab2}
\end{center} \begin{center}
\begin{tabular}{c|c|c|c|c|} \cline{2-5}
&\multicolumn{2}{|c|}{$|t|>|x|$}&
\multicolumn{2}{|c|}{$|t|<|x|$} \\ \cline{2-5}
&$(Rs),\,t>0$&$(Ls),\,t<0$&$(Us),\,x>0$&$(Ds),\,x<0$ \\ \hline
\multicolumn{1}{|c|}{$\cosh_e\th=$}&$\cosh \th$&$-\cosh \th$& $\sinh \th$&$-\sinh \th$ \\
\multicolumn{1}{|c|}{$\sinh_e\th=$}&$\sinh\th$&$-\sinh\th$&
$\cosh\th$&$-\cosh\th$ \\ \hline
\end{tabular} \end{center} \end{table}

The complete representation of the extended \hy functions can be obtained
by giving to $t,\,x$ all the values on the circle $t=\cos\f,\,x=\sin\f$
for $0\leq \f<2\pi$: in this way Eq. (\ref{sen1}) become:
\be \cosh_e\th=\frac{\cos\f}{\sqrt{|\cos 2\f|}}\equiv\frac{1}{\sqrt{|1-\tan^2\f|}};\;\;\;\;
\sinh_e\th=\frac{\sin\f}{\sqrt{|\cos 2\f|}}\equiv\frac{\tan\f}{\sqrt{|1-\tan^2\f|}}.\label{co22}
\ee These equations represent a bijective mapping between the points on unit
circle (specified by $\f$) and the points on unit hyperbolas (specified by $\th$).
From a geometrical point of view  Eq. (\ref{co22}) represent the projection,
from the coordinate axes origin, of the unit circle on the unit hyperbolas.
From the definitions of extended hyperbolic trigonometric functions of Eqs.
(\ref{cos}, \ref{sen}), we
can state for triangle in pseudo-Euclidean plane, exactly the same relations
between sides and angles as the ones that hold for Euclidean triangles  \cite{vin}. \\
\subsection{Trigonometry} \label{trigpe}
In fact let us consider a triangle in pseudo-Euclidean plane with no sides
parallel to axes bisectors: let us call $P_n\equiv(x_n,\,y_n)\; n=i,\,j,\,k
\,|\; i\neq j\neq k$ the vertices, $\th_n$ the \hy  angles.
The square \hy  length of the side opposite to vertex $P_i$ is defined
by Eq. (\ref{disq1}):
\be
D_i\equiv D_{j,\, k}=(z_j-z_k)(\tilde z_j-\tilde z_k) \mbox{ and } d_i=\sqrt{|D_i|}.
\label{disq3}
\ee
as pointed out before $D_i$ must be taken with its sign. \\
Following the conventions of Euclidean trigonometry we associate to the sides
three vectors oriented from $P_1\ra P_2;\;P_1\ra P_3 ;\;P_2\ra P_3$. \\
From (\ref{cos}), (\ref{sen}) and taking into account the sides orientation as done
in Euclidean trigonometry, we obtain:
\ba
\cosh_e \th_1 ={\displaystyle\frac{(x_2-x_1)(x_3-x_1)-(y_2-y_1)(y_3-y_1)}{d_2\,d_3}};\, 
\sinh_e \th_1 = {\displaystyle\frac{(x_2-x_1)(y_3-y_1)-(y_2-y_1)(x_3-x_1)}{d_2\,d_3}}\nonumber\\
\cosh_e \th_2 ={\displaystyle\frac{(y_3-y_2)(y_2-y_1)-(x_3-x_2)(x_2-x_1)}{d_1\,d_2}};\,
\sinh_e \th_2 ={\displaystyle\frac{(x_2-x_1)(y_3-y_2)-(y_2-y_1)(x_3-x_2)}{d_1\,d_2}} \nonumber\\
\cosh_e \th_3 ={\displaystyle\frac{(x_3-x_2)(x_3-x_1)-(y_3-y_2)(y_3-y_1)}{d_1\,d_3}};\,
\sinh_e \th_3 ={\displaystyle\frac{(x_3-x_1)(y_3-y_2)-(y_3-y_1)(x_3-x_2)}{d_1\,d_3}} \nonumber
\label{ele}
\ea
It is straightforward  to verify
that all the functions $\sinh_e\th_n$ have the same numerator. If we call this
numerator:
\be
x_1(y_2-y_3)+x_2(y_3-y_1)+x_3(y_1-y_2)=2S \label{Sup}
\ee
we can write:
\be
2S=d_2\,d_3\sinh_e\th_1=
d_1\,d_3\sinh_e\th_2=d_1\,d_2\sinh_e\th_3 \label{area}.
\ee
In Euclidean geometry a quantity equivalent to $S$ represents the area of the triangle.
In pseudo-Euclidean geometry $S$ is still {\it an
invariant quantity linked to the triangle}. For this reason it is
appropriate to call $S$ {\it pseudo-Euclidean area} \cite{Ya2}. \\
We note that the expression of area (Eq. \ref{Sup}), in terms of vertices
coordinates, is exactly the same as in Euclidean geometry (Gauss formula
for a polygon area applied to a triangle). \\


\begin{thebibliography}{15}
\bibitem{nota2} The value of $\cosh_e$ must be interpreted following table 1
in Appendix \ref{inv}. 
\bibitem{nota3} More precisely only sides directions parallel to axes bisectors
have to be excluded.
\bibitem{nota1} From a physical point of view the speed can not change in a null
time, but this time can be considered short with respect to $\t_1,\,
\t_2$ (\cite{Na}, pag. 41).\\ An experimental result with just uniform motion is
reported in \cite{CERN}. In this experiment the lifetime of the muon in the CERN
muon storage ring was measured.
\bibitem{noi} In \cite{cz} the hyperbolic motion is obtained as a straightforward
consequence of the invariance of the wave equation with respect to Lorentz transformations.
\bibitem{notaper} We recall that two straight-line are
pseudo-orthogonal if they are symmetric with respect to a couple of lines parallel
to the axes bisectors \cite{vin, Na}.
\bibitem{nota27}The following equation has an explicit solution for $\a=2$. In fact we have \\
$ \cosh( \th_2-\th_1)=2\,\sinh^2\th_1+1\equiv \cosh 2\,\th_1 \Ra \th_2=3\,\th_1 .$
\bibitem{nota28} Since the total time is the same, shorter time with uniform motion
means longer time with accelerated motion, so this result is that of Eq. (\ref{ratio2})
\bibitem{notainv} Usually just one invariant quantity given by the scalar product is
considered \cite{Fj2, Na}. The introduction of a second one allows us to consider
Euclidean trigonometry in a new way and to formalize space-time trigonometry
\cite{vin} as summarized in appendix \ref{inv}.
\end{thebibliography}
\end{document}